\documentclass[12pt]{article}\usepackage[hyperfootnotes=false]{hyperref}
\usepackage{epsfig}
\usepackage{float}

\usepackage{caption}

\usepackage{amsmath}
\usepackage{amssymb}
\usepackage{graphicx}
\newlength{\abstractwidth}
\renewcommand{\thefootnote}{\fnsymbol{footnote}}
\renewcommand{\thanks}[1]{\footnote{#1}} 
\newcommand{\starttext}{
\setcounter{footnote}{0}
\renewcommand{\thefootnote}{\arabic{footnote}}}
\renewcommand{\theequation}{\thesection.\arabic{equation}}
\newcommand{\be}{\begin{equation}}
\newcommand{\bea}{\begin{eqnarray}}
\newcommand{\eea}{\end{eqnarray}}
\newcommand{\beq}{\begin{equation}}
\newcommand{\ee}{\end{equation}}
\newcommand{\eeq}{\end{equation}}

\def\ba{\begin{eqnarray}}
\def\ea{\end{eqnarray}}

\def\12{{1 \over 2}}
\def\eq{&=&}

\def\la{\langle}
\def\ra{\rangle}

\def\simleq{\; \raise0.3ex\hbox{$<$\kern-0.75em
\raise-1.1ex\hbox{$\sim$}}\; }
\def\simgeq{\; \raise0.3ex\hbox{$>$\kern-0.75em
\raise-1.1ex\hbox{$\sim$}}\; }

\def\O2{\Omega_2}

\def\bi{\begin{itemize}}
\def\ei{\end{itemize}}

\def\sc{\setcounter{equation}{0}}

\def\W{$\Omega$}
\def\W'{$\Omega$}

\def\V{\Omega}
\def\V'{\Omega}

\def\O{${\cal{O}}$}

\def\c{{\cal{C}}}

\def\bn{\bigskip \noindent}

\def\suk{SU(2^K)}

    \def\cg{$\c$-geometry}
     \def\cg2{$\c_2$-geometry}

\makeatletter
\g@addto@macro\normalsize{%
  \setlength\abovedisplayskip{10pt}
  \setlength\belowdisplayskip{20pt}
  \setlength\abovedisplayshortskip{10pt}
  \setlength\belowdisplayshortskip{20pt}
}
\makeatother

\usepackage{color}

\begin{document}
\renewcommand{\theequation}{\thesection.\arabic{equation}}
\begin{titlepage}
\rightline{}
\bigskip
\bigskip\bigskip\bigskip\bigskip
\bigskip
\centerline{\Large \bf { ER=EPR, GHZ,  and the Consistency}}
\centerline{\Large \bf { of Quantum Measurements}}

\bn

\bigskip
\begin{center}
\bf  Leonard Susskind  \rm

\bigskip

Stanford Institute for Theoretical Physics and Department of Physics, \\
Stanford University,
Stanford, CA 94305-4060, USA \\
\bigskip

\end{center}

\begin{abstract}
This paper illustrates various aspects of the ER=EPR conjecture.
It begins with a brief heuristic argument,  using the Ryu-Takayanagi correspondence, for why  entanglement between black holes implies the existence of  Einstein-Rosen bridges.

The main part of the paper addresses a fundamental question:  Is ER=EPR consistent with the standard postulates of quantum mechanics? Naively it seems to lead to an inconsistency between observations made on entangled systems by different observers.
The resolution of the paradox  lies in the properties of multiple black holes, entangled in the Greenberger-Horne-Zeilinger pattern.

The last part of the paper  is about  entanglement  as a resource for quantum communication. ER=EPR provides a way to
visualize protocols like quantum teleportation. In some sense  teleportation takes place through the wormhole, but as usual, classical communication is necessary to complete the protocol.

\medskip
\noindent
\end{abstract}


\end{titlepage}

\starttext \baselineskip=17.63pt \setcounter{footnote}{0}

\sc
\section{ER=EPR}\label{S-ER=EPR}

The AMPS paradox  \cite{Almheiri:2012rt}  leads to one of two conclusions: Either there are firewalls at the horizons of black holes that are highly entangled with other systems, or entanglement implies the kind of geometric connectivity described by ER=EPR \cite{Maldacena:2013xja}.
Firewalls provide a relatively straightforward resolution of the paradox, which doesn't require us to rethink quantum mechanics. It does require a   mechanism  for firewall formation which is absent at present.

By contrast ER=EPR  touches on the deepest  foundations of quantum mechanics  in ways that may profoundly influence the future interpretation of the subject. The main part of this paper is about the relation between Einstein-Rosen bridges and the consistency of different observers' descriptions of reality. More specifically  we will take up the question of what happens to an Einstein-Rosen bridge, when a complete set of measurements at one end  collapses the entangled wave function into an unentangled product.

 I will start with a short argument in support of ER=EPR.
By now it is clear that the existence of an ERB  between two black holes implies that they are entangled. In other words $ER \Rightarrow EPR.$ The converse---$EPR\Rightarrow ER$---states that entanglement between two black holes implies that they are connected by an ERB. This latter statement is considered less certain. What follows in this section is not meant to be a proof---it has an element of circularity to it---but combining ER=EPR with RT provides a neat package which lends  support for $EPR\Rightarrow ER.$

\begin{figure}[H]
\begin{center}
\includegraphics[scale=.3]{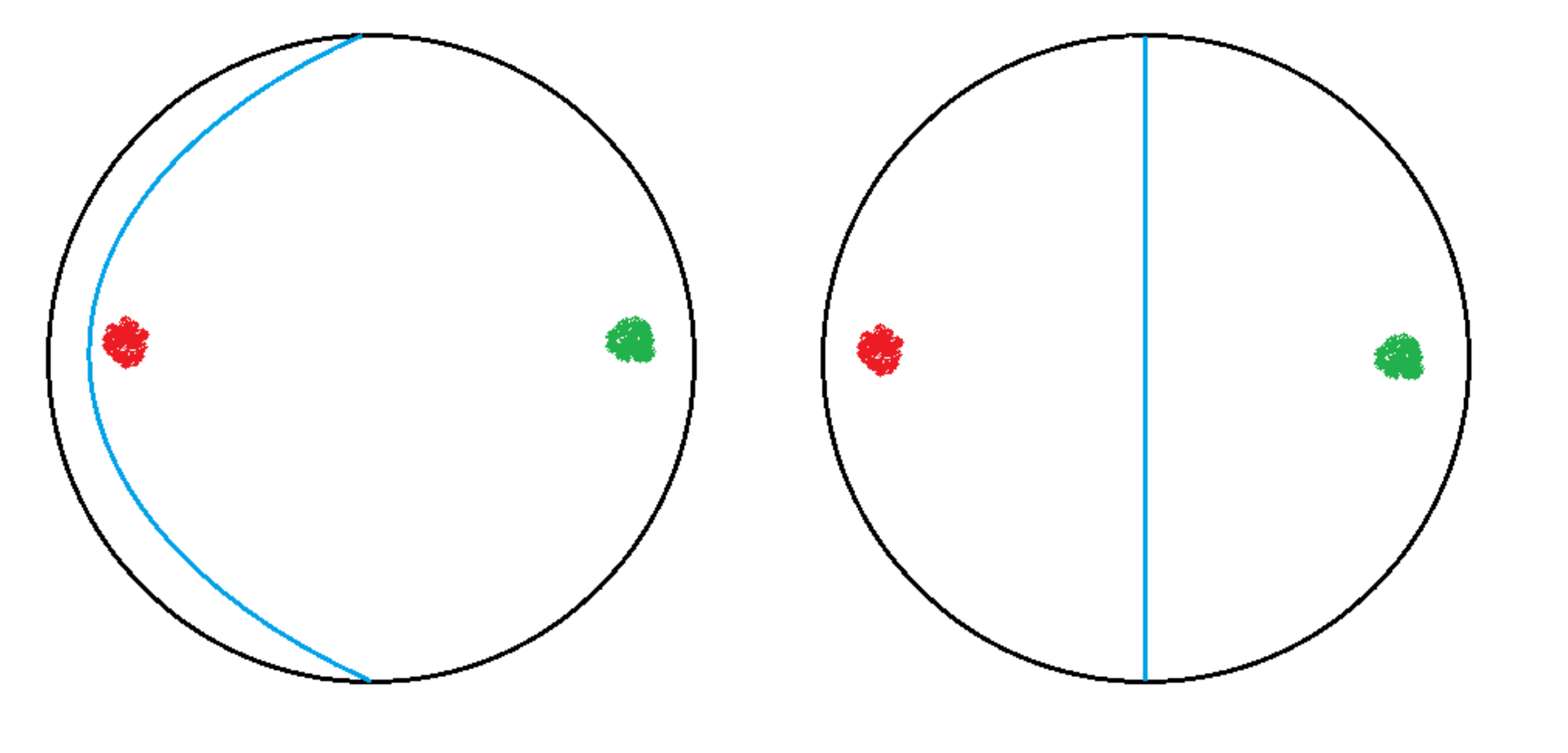}
\caption{A special slice of ADS with two objects. The blue line represents the RT surface for calculating entanglement entropy
between the left and right semicircles. The construction fails to account for possible entanglement between the objects.}
\label{RT}
\end{center}
\end{figure}

Consider a spatial slice of ADS containing two objects in separate non-overlapping regions of space.
The objects are placed near the boundary at diametrically opposed positions.
In figure \ref{RT} they are shown as  red and green blobs. I will assume that the configuration is instantaneously static. This can be ensured by making the boundary CFT wave function  time-reversal symmetric. For simplicity  I will illustrate the case  $(2+1)$-dimensional ADS.

Let us consider the entanglement entropy of the two complementary boundary semicircles on the left and right. According to the Ryu-Takayanagi correspondence\cite{Ryu:2006bv} one should consider curves which are homologous to either  of the semicircles. The length of the shortest such curve in Planck units gives the entanglement entropy. If the two objects are near the boundary the shortest geodesic will be the vertical blue line and its length will be almost the same as the pure ADS case. That's fine as long as the two objects are not entangled, but if they are,  the entanglement entropy should be greater  by the entanglement entropy of the objects \cite{Faulkner:2013ana}\cite{Engelhardt:2014gca}. Naively there does not seem to be any way for the unmodified RT formula to account for this difference. In what follows I will explain how ER=EPR allows us to expand the range of validity of RT, and account for additional entanglement.

Let's replace the two objects in figure \ref{RT} by black holes. If the black holes are unentangled then the global topology of the spacelike slices is unchanged by the presence of the black holes  \cite{Susskind:2014moa}. They each contain a ``bridge-to-nowhere" \cite{Susskind:2014jwa}  but that does not obstruct the smooth deformation of the semicircle to the short vertical geodesic. Thus the unentangled black holes do not change the entanglement entropy of the two sides.

\begin{figure}[H]
\begin{center}
\includegraphics[scale=.3]{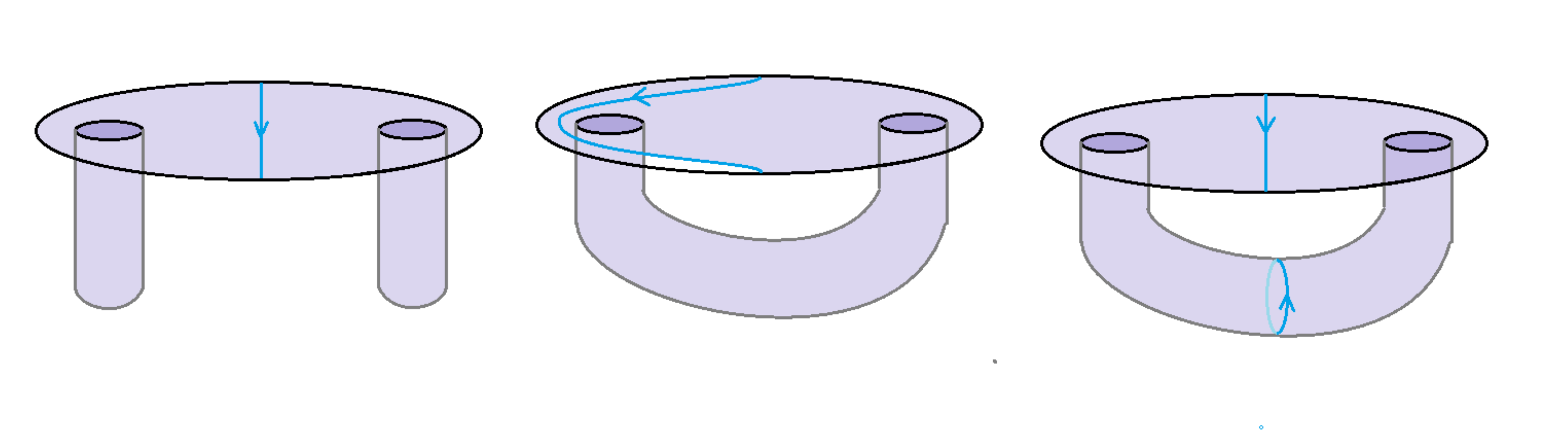}
\caption{The RT surface for two black holes contained in ADS. If the black holes are not entangled  then the topology is trivial as in the left panel. If they are
entangled the validity of the RT prescription requires a wormhole between them.This is shown in the middle and right panels.}
\label{BH}
\end{center}
\end{figure}

But if the black holes are entangled, say in the TFD state, they should add an entanglement entropy equal that of a single black hole. This must mean that there is an obstruction to naively pulling the RT surface to the line between the black holes.  The resolution is that the entanglement creates an ERB as in figure \ref{BH}.

 In pulling the RT curve past the black hole a piece gets caught on the wormhole as in the last of the pictures. The result is that the shortest homologous RT path consists of two pieces. The first is the vertical blue line which gives the same answer as the unentangled black holes. The second piece adds the entanglement of the two black holes. Thus as we expect, a pair of entangled objects adds to the entanglement of the vacuum. Turning it around, the requirement that the entangled objects contribute their entanglement entropy requires the existence of the ERB.

 We can generalize this to a system of small black holes.
 \begin{figure}[H]
\begin{center}
\includegraphics[scale=.3]{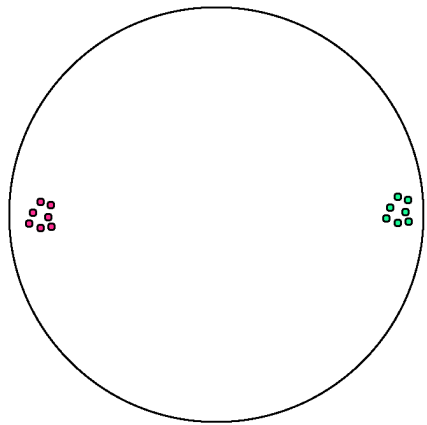}
\caption{The entangled systems are clouds gas composed of small black holes. }
\label{gas}
\end{center}
\end{figure}
 Suppose there are $N$ very small black holes on one side, entangled or not entangled  with $N$ similar black holes on the other side. Let's think of them as two  clouds of gas whose atomic constituents are the small black holes. There are two ways we can proceed. The first is phenomenological: begin by constructing a phenomenological equation of state for the gas and an energy density, pressure, etc.  Once this is done we can feed these into the Einstein equations and calculate the effects on the geometry. This procedure will not depend on whether the red and green black holes are entangled. If we use the corrected geometry as input to RT we will not be able to distinguish the cases. Thus we would have to add a phenomenological term to the RT prescription. This has  been elegantly formulated in \cite{Faulkner:2013ana}\cite{Engelhardt:2014gca}.

 The other way to proceed is microscopic. If the black holes are entangled then ER=EPR implies wormholes.
 The minimal RT surface will contain the vertical blue component and also $N$ small loops winding around the ERBs. Without those loops there would be no difference between the entangled and unentangled case.

 What about more general types of blobs? For example how should we argue if the black holes are replaced by localized clouds of elementary particles? The phenomenological way of adding a contribution for the entangled clouds is certainly possible.  It's also tempting to imagine that the entangled particles are connected by highly quantum versions of ERB's but that is hard to make precise.   I will give a different perspective.

 The answer that I propose is that entanglement is a resource (see section \ref{S-resource}) that can come in many forms. Entangled photons is one form, Einstein-Rosen bridges connecting black holes is another. Moreover the resource is fungible. That means that the various forms can be converted into one another. For example two extremely distant entangled black holes can evaporate and turn into two entangled clouds of quanta. Or two entangled clouds can be collapsed to produce a black hole pair connected by an ERB.  The laws of quantum mechanics allow conversion from one form to another by local operations as long as the entanglement entropy is conserved.

 Now let's consider the entangled blobs in figure \ref{RT}. If they are not themselves black holes,  without changing the entanglement entropy we can collapse them into black holes. Then we can use the geometric argument and calculate the entanglement entropy using RT as in figure \ref{BH}.

By this argument we can extend the range of validity of the RT conjecture to configurations of large but fixed entanglement entropy.

\sc
\section{Preliminary Notes}\label{preliminary}

Before beginning a discussion of measurements and ERBs
let's quickly review some  facts and assumptions.
\subsection{Note about Measurements }\label{S-Measurement}

\bn

There are two ways to think about measurements. In the first way we simply accept the projection or collapse postulate. The result of a measurement is  recorded, and the state is  projected  onto a simultaneous eigenstate of the measured observables.
Let $I$ index the eigenvalues of some complete set of commuting observables for the system of interest. The wave function in the $I$ basis is $\Psi(I) = \la I|\Psi\ra.$ The Born rule gives the probability

\be
P(I) =    \Psi^{\ast}(I) \Psi(I)
\ee
for the outcome $I.$

Sometimes this is repackaged by assembling the various outcomes into a density matrix,

\be
\rho = \sum_I P(I) |I\ra \la I |
\ee
Note that we have lost track of the phases of the $\Psi(I).$  In this view a measurement is irreversible.

In the second way,  a measurement is a process that  entangles the system with the apparatus. Before the measurement took place the system and  apparatus were in an unentangled product state. After the measurement they will be entangled. If we label the orthonormal  pointer states of the apparatus as $|\phi(I)\ra$ and the initial apparatus state by $|\phi_0\ra,$ then the measurement is described (unitarily) by,

\be
\sum_I \Psi(I) \ |I\ra \otimes |\phi_0\ra  \to \sum_I  \Psi(I) \ |I\ra\otimes |\phi(I)\ra
\ee

The entangled state maintains information about the phases of $\Psi(I),$ which in principle can play a role if the measurement is undone (See later discussion). In most of this paper we will think of measurements as the establishment of entanglement. In the last section on teleportation the simpler projection postulate will be good enough for our purpose.

\subsection{Note about Qubits}\label{S-qubits}

The assumption that the degrees of freedom of a black hole can be represented as a system of qubits is probably not very restrictive. A more restrictive assumption that was first spelled out by Hayden and Preskill \cite{Hayden:2007cs} is that there is a preferred qubit basis in which the Hamiltonian has a special property called $n$-locality. The assumption of $n$-locality is quite simple. It says that the Hamiltonian is built from a sum of terms, each of which is a product of no more than $n$ qubit operators. Moreover $n$ is assumed to be a small number that does not grow as the total number of qubits grows.

 Most lattice Hamiltonians such as Ising and Heisenberg magnets, are 2-local since they are sums of 2-spin operators. Single spin operators can be added without changing the 2-locality. These Hamiltonians are not only $n$-local but they are also spatially local. However $n$-locality does not require any sense of spatial locality. In general there can be terms involving any combination of $n$ or fewer qubits. $n$-locality does involve a choice of basis for the qubits. A general unitary operator acting on the Hilbert space will map the qubits to a new set of qubits in which the Hamiltonian will not be $n$-local. I will call a set of qubits, in which $H$ is $n$-local, computational qubits. We can further fix the basis by diagonalizing the the  $Z$-components of all the computational qubits to define a computational basis.

 The main reason for believing that the Hamiltonian is $n$-local  is the fast-scrambling property of black holes. I will not go through the argument here but just refer to the literature on fast-scrambling \cite{Hayden:2007cs}\cite{Sekino:2008he}\cite{Lashkari:2011yi}.

\subsection{Note about Precursors}
Two observers who happen to share a bipartite entangled system with a very large entanglement entropy can in principle meet after a time which is much shorter than the distance between them. A well known example is two black holes entangled in the thermofield double state of a pair of  CFTs. This is an extreme case in which the black holes can be considered to be an infinite distance apart, since the two CFTs are completely non-interacting.

\begin{figure}[H]
\begin{center}
\includegraphics[scale=.3]{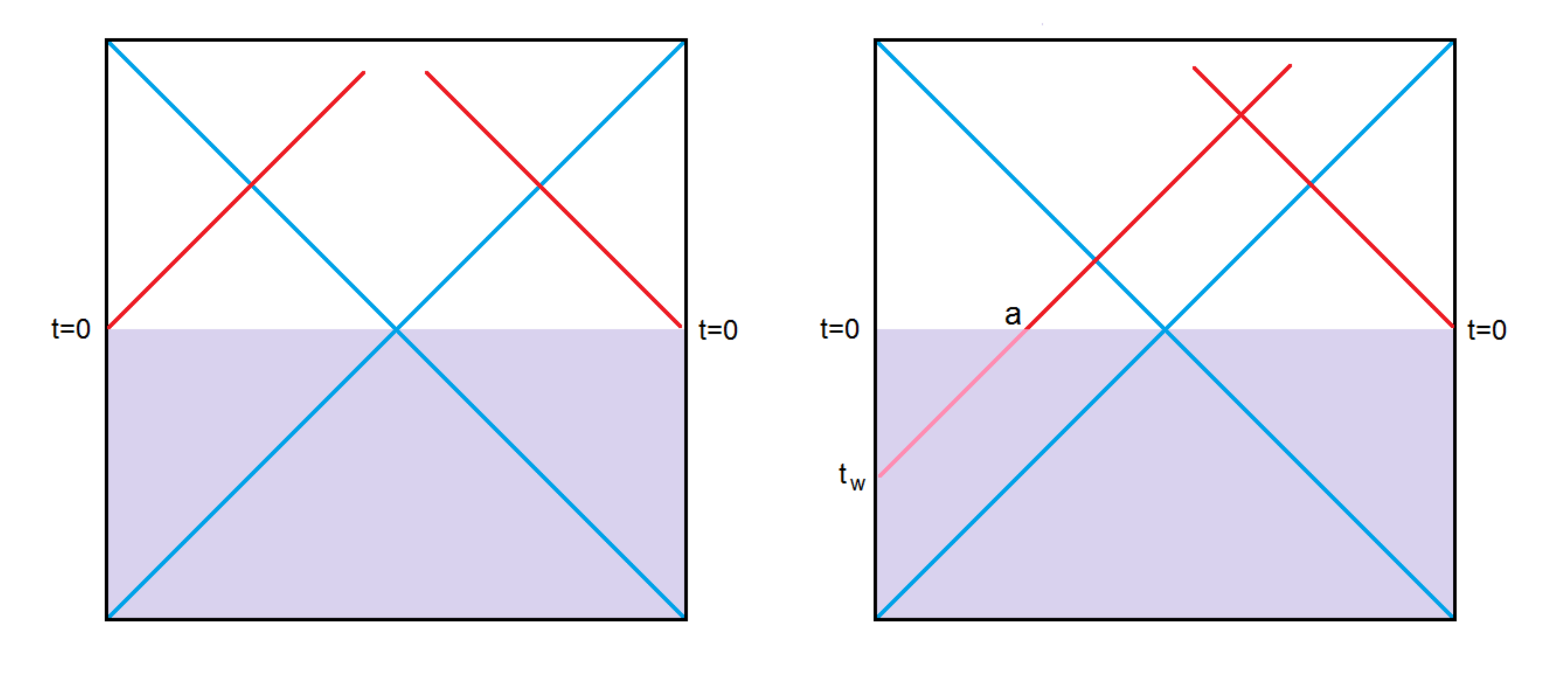}
\caption{On the left side the signals originate at the boundary at $t=0.$ They barely meet before they reach the singularity. In the right figure Charlie creates a signal by acting with a precursor which effectively introduces the signal at point $a.$ No interaction between Charlie's side and Bob's side is necessary in order for their signals to meet and interact in the ERB.}
\label{meet}
\end{center}
\end{figure}

Let us suppose Charlie and Bob each control one of the CFTs but there are no interactions between them. Charlie and Bob cannot send messages as long as they stay outside the black holes, but signals that they send into the black holes can meet
 in the interior of the Einstein-Rosen bridge. The signals could be Charlie or Bob themselves. Figure \ref{meet} shows how this happens. We assume that the black holes are created at $t=0$ in the TFD state. The red lines represent the trajectories of Charlie's and Bob's signals. Note that if they enter the geometry from the boundary later than $t=0$ they will not meet.

However, if we assume that Charlie and Bob have sufficient control over their own CFTs, then they can send  signals by acting with precursors \cite{Susskind:2014rva}. One may think of the precursors as non-local operators that reach into the bulk from a point labeled $a$ in the right side of figure \ref{meet}.

If we assume that Charlie and Bob have sufficient control then it is possible for them to  convert their entangled systems from one form to another as long as the entanglement entropy does not increase. In fact if they know the state of the bipartite system they can transform it to the thermofield double state by strictly local operations, each observer acting only at his own end. In particular quantum mechanics presents no obstacle to them converting their entangled systems to a pair of black holes in the TFD state.

Thus to put is simply, given a large enough entanglement resource, Charlie and Bob can convert it to a TDF state of two black holes, and then jump in and meet in the ERB.

\subsection{Note about Difficult Experiments}

Some of the thought experiments that follow may seem so difficult that one might question whether they can be carried out even in principle. For example in section \ref{S-Measurement and ERBs} Alice will be asked to make a complete set of commuting measurements on the degrees of freedom of Bob's black hole (which is entangled with Charlie's black hole).

Making such a measurement  may seem completely unfeasible, but for our  purposes  it's not  necessary. An important theme of this paper is that from an information theoretic standpoint a black hole is equivalent to the matter that formed it. For example the thermofield-double state can be constructed by starting with a system of Bell pairs divided into Bob's and Charlie's shares, which are subsequently collapsed into black holes. Alice's measurement may be carried out \it before  \rm the qubits are collapsed. In this way of thinking, the measurement produces a system of $ghz$ triplets which are later converted into black holes. A similar strategy for evading the Harlow-Hayden obstruction was described by Oppenheim and Unruh  \cite{Oppenheim:2014gfa}.

Some of the thought experiments involve operations which shorten an Einstein-Rosen bridge. These effectively require putting a black hole into a machine which unitarily reverses the time evolution of the black hole. This does not violate any law of quantum mechanics but one can substitute a different strategy, at least if evaporation is not an issue. Instead of acting to shorten the ERB one can wait for nature to do the job. The quantum recurrence theorem guarantees that in a predictable time the Hamiltonian evolution of the isolated black hole will shorten the ERB.

\sc
\section{Measurements and Einstein-Rosen Bridges}\label{S-Measurement and ERBs}

Three  things  distinguish quantum physics from classical physics; entanglement, the capacity for exponential complexity, and the collapse postulate when a measurement is made\footnote{This is also called the projection postulate.}. These are not unrelated: the process of measurement is understood as the establishment of entanglement between a system and an apparatus: the irreversibility of a measurement is ensured by the subsequent onset of complexity in the apparatus (or the environment): and  collapse is the  way that a single observer describes its own measurement. But there is not just one observer in the world and the observations of different observers must be knitted together in a consistent way \cite{Everett}\cite{Riedel:2013uoa}. An important part of this has to do with the fact that when an observation is made, the result is recorded so that other observers can consult the record. Standard quantum mechanics guarantees consistency  in a remarkable way that does not leave room for modifications.

In order to approach the problem of the consistency of ER=EPR with measurement theory  I will consider a particular situation involving a pair of black holes and three observers; Alice, Bob, and Charlie. Later we will add Daisy.
The two black holes will be assumed to be far apart. One will be under the control of Bob and the other under the control of Charlie. 

We will assume that each observer possesses an arbitrarily powerful machine which  can be used to carry out any unitary transformation,  or to make arbitrary measurements, on any system in its own vicinity. In particular the observers can apply their machines to black holes so as to transform their states, or to make measurements on them. Unless otherwise specified, they will only be allowed to use their machines locally. 

We will assume that Bob's and Charlie's black holes are maximally entangled with one another, and are  therefore  connected by an ERB. We further assume that they can send signals which can meet in the ERB. Bob and Charlie may even jump in themselves and meet. This typically will involve computationally complex operations  (precursors)  but  with the help of their machines  the operations can  be carried out \cite{Maldacena:2013xja}\cite{Susskind:2014rva}.
 Finally we  assume that a black hole can be described in terms of a collection of $K$ ``computational" qubits where $K$ is of order the entropy.

Now let us suppose that Alice comes to Bob's black hole and makes a measurement of a complete set of $K$ commuting observables. For example she can measure all the $Z$  components of the computational qubits comprising Bob's black hole. In order to record the measurement Alice must be in possession of a memory system which itself must contain at least $K$ qubits. Once the measurement has been made Alice will say that Bob's black hole is in a pure state; namely one of the computational eigenstates. Since Bob's and Charlie's black holes were originally entangled she  also says that Charlie's black hole is in a similar pure state and that the two are not entangled. Alice must conclude that there is no ERB connecting the black holes. She may warn Bob and Charlie that no matter what they do with their machines (locally) they cannot meet in the ERB.

Now let's introduce Alice's friend Daisy who knows about the entire setup, but has not interacted with it since the start of the experiment.  Daisy has also not been in any contact with the other observers. How does she describe things?   Daisy's description is that the   system composed of the  two black holes, together with Alice's memory,  are are in an entangled tripartite state. One can go further and allow Alice to squeeze her memory into a black hole so that the tripartite system consists of three black holes. According to Daisy the three systems are entangled, so there should be some kind of ERB connecting them. One might expect that Bob and Charlie can meet in that ERB. On the face of it there seems to be a contradiction between Alice's and  Daisy's conclusions.

To address this paradox let me review what is known about multipartite entanglement of several black holes.
A class of multiple boundary ERBs in ADS have been studied in \cite{Gharibyan:2013aha}\cite{Balasubramanian:2014hda}. By assumption the interior geometries of these ERBs are smooth, i.e., they are described by classical geometry. I will assume that there is no obstruction to various combinations of parties, or their signals,  meeting inside.

If this were the type of system created by Alice's measurement there would indeed be a contradiction between Alice's view (Bob and Charlie cannot meet) and  Daisy's (yes they can). In fact we will see that the tripartite state created by the measurement is very different from those described  in \cite{Gharibyan:2013aha}\cite{Balasubramanian:2014hda}.

One of the properties of the ERBs studied in \cite{Gharibyan:2013aha}\cite{Balasubramanian:2014hda} is that they have little or no $GHZ$ entanglement. This does not mean that $GHZ$ entanglement can never occur or that it has no interesting role to play. What it does mean is the $GHZ$-entanglement introduces non-classical features in the ERB. We will see that this type of entanglement is intimately  involved in the measurement process described above. I will explain by starting with the
 simplest case in which the black holes and Alice's measuring apparatus are all replaced by single qubits.
  A basis of state vectors for the three qubit system can be written as,

\be
| c, b, a \ra
\ee
where $a, b,$ and $c$ all take on the values $(0,1).$  Here $a$ refers to Alice's apparatus qubit, and $b,c$ to the two qubits comprising the entangled system. The apparatus interacts with the qubit $b$ and measures it. The measurement is described by a unitary operator that has the action,

\bea
|c,0,0\ra &\to& |c,0,0\ra \cr \cr
|c,1,0\ra &\to& |c,1,1\ra \cr \cr
|c,0,1\ra &\to& |c,0,1\ra \cr \cr
|c,1,1\ra &\to& |c,1,0\ra
\eea

In the initial state the apparatus is in the state $|0\ra$ and the entangled system is in the state $u|0,0\ra + v|1,1\ra.$ I will assume that $u$ and $v$ are almost equal so that the state is close to maximally entangled\footnote{The only reason for not making $u$ and $v$ equal is in order to resolve an ambiguity about which basis Alice's measurement takes place in. }.
The measurement is summarized by,

\be
u|0,0,0\ra + v|1,1,0\ra \to u|0,0,0\ra + v|1,1,1\ra
\label{uv}
\ee
If $u=v$ the result of the measurement is the tripartite $ghz$-state\footnote{I use lower case $ghz$ for the state of three qubits. Upper case $GHZ$ will be used for analogous states of black holes.}. This is the simplest example of the fact that whenever a measurement is done on one member of an entangled system, some degree of $ghz$-entanglement is induced.

 The $ghz$-state can be expressed as a tensor $T_{ijk}$ with three indecies representing the three qubits. Expressed symmetrically, the tensor $T_{ijk}$ is defined by,

 \bea
 T_{ijk} \eq  \delta_{ij}\delta_{jk}\delta_{ki}   \ \ \ \ \ ( \rm no \ sum ) \cr \cr
  |ghz\ra \eq  T_{ijk}|i,j,k\ra  \ \ \ \ \ ( \rm yes \ sum )
 \eea

A more general state with $ghz$-entanglement can be described by transforming $T$ by local unitary operators,

 \be
 T^{\prime}_{i^{\prime}j^{\prime}k^{\prime}} =  U_{i^{\prime}i}  V_{j^{\prime}j}  W_{k^{\prime}k}  T_{ijk}
 \ee

 States with $ghz$-entanglement have the following properties:
\begin{enumerate}
\item If any two qubits are traced over, the density matrix of the third qubit is  maximally mixed. Therefore any qubit is  maximally entangled with the union of the other two.
\item If any one qubit is traced over the density matrix of the other two is separable. For example in the case of the state \ref{uv} the two-qubit density matrix is

\be
\rho =|u|^2 |0,0\ra \la 0,0| + |v|^2 |1,1\ra \la 1,1|
\label{separable}
\ee
A separable density matrix like \ref{separable} is distinguished by having no entanglement between any two parties. They are of course classically correlated.
\end{enumerate}

Now let us come to the black hole case. Assume that we have a pair of entangled black holes, labeled $B,C$  which for definiteness are in the thermofield-double state. For simplicity I will replace the TFD state by a maximally entangled state of $2K$ qubits.
Let $I$ label a set of orthonormal basis states of $K$ qubits. The initial state of the black hole pair is given by,

\be
|TFD\ra = \sum_{I=1}^{2^K}|I,I\ra
\label{TFD1}
\ee

The state can also be written as a product of $K$ Bell pairs,

\be
|TFD\ra = \left\{   |00\ra +|11\ra      \right\}^{\otimes K}
\label{TFD2}
\ee

It is worth noting that the states \ref{TFD1} and \ref{TFD2} are basis-independent under any local unitary transformation\footnote{By a local unitary transformation I mean one which is a product of unitary operators each acting on one black hole.}.

Alice wishes to make a measurement of a complete commuting set of observables of black hole $B.$ In order to record the results of the measurement her apparatus must have at least $K$ qubits. Therefore we have a tripartite system of $3K$ qubits. Initially the black hole pair $B,C$ is in the state \ref{TFD1} or equivalently \ref{TFD2} and Alice's apparatus is in the state $|000000....\ra.$
After the measurement the overall state is,

\be
|GHZ\ra = \sum_{I=1}^{2^K}|I,I,I\ra
 \label{after measurement}
\ee

Unlike the TFD state in \ref{TFD1} the state \ref{after measurement} is basis dependent under local unitary transformations. Different choices of the basis $|I\ra$ lead to physically different states, the tripartite ERB being highly dependent on the basis. However, the different choices are within the same equivalence class under local unitary transformations of the three parties.

If the basis is a qubit basis  in which each qubit is in the state $|0\ra$ or $|1\ra$ then \ref{after measurement} can be written in the form

\bea
|GHZ\ra \eq \left\{   |000\ra +|111\ra      \right\}^{\otimes K} \cr \cr
\eq |ghz\ra^{\otimes K}
\label{GHZ}
\eea
In other words it is a product of $ghz$ triplets distributed among $C,B$ and $A.$

\begin{figure}[H]
\begin{center}
\includegraphics[scale=.3]{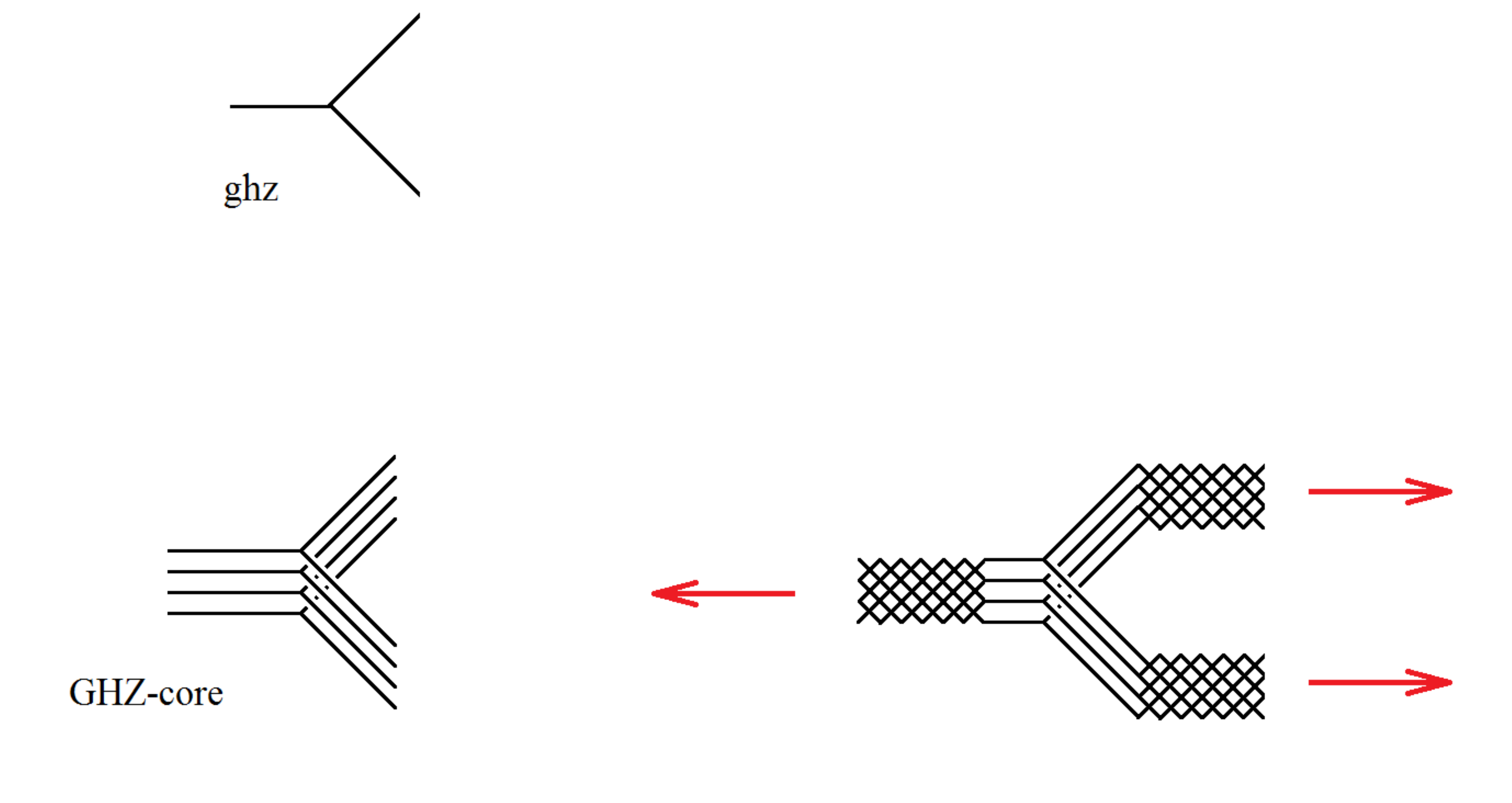}
\caption{Tensor network representation of $GHZ$-entangled black holes. The upper figure shows the tensor $T_{ijk}$ representing the basic $ghz$-triplet. The lower left shows  tensor network for a product of  $ghz$-triplets. The lower right illustrates the evolution of the tensor network with time. The growth of the tensor network represents the growth of complexity.}
\label{ghz}
\end{center}
\end{figure}

Let us assume that the qubit basis in \ref{GHZ} is the computational basis in which the Hamiltonian is $n$-local.  A good way to describe the ERB and its evolution is by constructing a tensor network. Figure \ref{ghz} illustrates the construction.
The first part of the figure shows the basic vertex representing three qubits in a $ghz$ state. The tensor in the qubit basis is zero unless all three edges are the same, in which case the value of the tensor is $1.$  The product state in \ref{GHZ} is shown next as a simple juxtaposition of $K$ elementary $ghz$ vertices.

That's not the end of the story; the state evolves toward increasing complexity according to the rules of \cite{Susskind:2014rva}\cite{Stanford:2014jda}\cite{Roberts:2014isa}. The evolution, shown in the lower right of figure \ref{ghz}, adds layers to each end of the ERB by growing the tensor network outward from the central $GHZ$ core  \cite{Hartman:2013qma}.

The tensor network in figure \ref{ghz} suggests that the ERB connecting $A,$ $B,$ and $C$  will evolve to something smooth and classical throughout most of its volume, the non-classical $GHZ$ core being localized  in a receding region far from the horizons. Ordinarily it would not affect an infalling observer who crosses the horizon  more than a Schwarzschild time (or an ADS time) after the measurement.  Later, however, if we allow a machine to shorten the ERB at one end (in general a computationally very complex operation)  then it could be possible for an in-falling observer  to experience something abnormal.

\begin{figure}[H]
\begin{center}
\includegraphics[scale=.3]{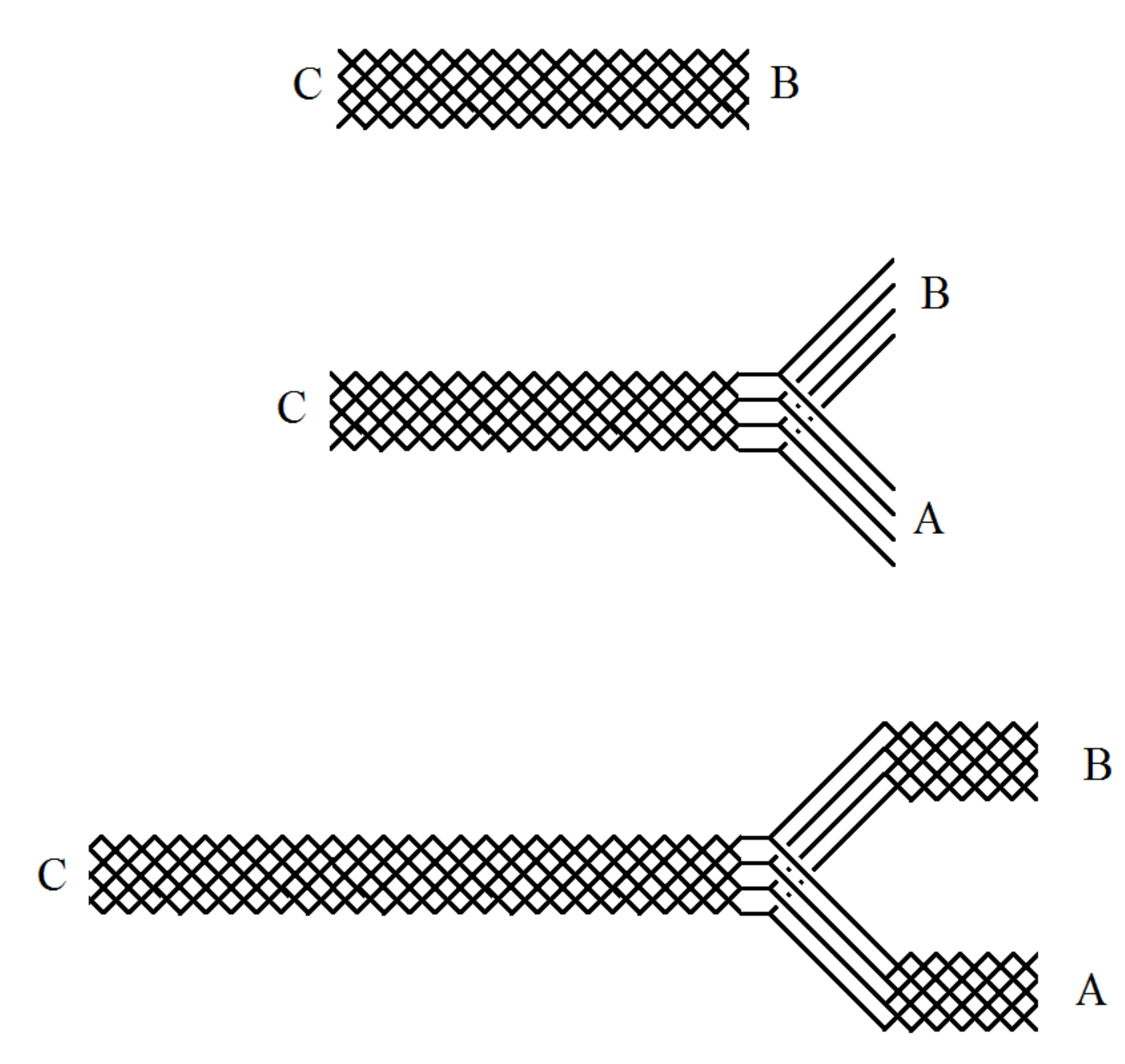}
\caption{Instead of  the TFD state we can start with an entangled pair of black holes which has already evolved to a complex state. The top panel shows the tensor network for the starting point. The middle panel shows the tensor network right after Alice makes a measurement. The bottom panel shows the subsequent evolution toward increasing complexity. }
\label{EVO}
\end{center}
\end{figure}

We can consider a more general situation in which the starting point is a time-evolved TFD state represented by a long tensor network \cite{Hartman:2013qma}\cite{Stanford:2014jda}\cite{Roberts:2014isa}\cite{Susskind:2014moa}. In figure \ref{EVO} the evolution starts at the top panel with the Bob-Charlie entangled black holes. The initial state \ref{TFD1} is replaced by,

\be
\sum_{IJ} U_{IJ} |I,J\ra
\ee
where $U_{IJ}$ is a unitary matrix representing the action of the tensor network in the top panel.

In the second panel Alice has just made her measurement. Subsequently all three subsystems evolve as in the bottom panel.

\bn

The properties of  $GHZ$ states parallel the properties of the simpler $ghz$ state of three qubits:

\begin{enumerate}
\item If any two black holes are traced over, the density matrix of the third black hole is  maximally mixed. Therefore any of the three black holes is  maximally entangled with the union of the other two.
\item If any one black hole is  traced over the density matrix of the other two  is separable so that no two black holes are   entangled. They are  classically correlated.
\end{enumerate}

The tripartite ERBs represented in figures \ref{ghz} and \ref{EVO} consist of three  legs knotted together at the center by a symmetric core representing the $GHZ$-state.  The three legs of the tensor network eventually end at the three horizons and are smooth from the horizons to the $GHZ$-core which cannot be removed by any local unitary operations since $GHZ$-entanglement is invariant under such transformations. The resolution of the apparent contradiction described earlier lies in the properties of the $GHZ$-core which must obstruct Bob and Charlie from meeting in the ERB.

In fact from each of their points of view the $GHZ$-core  behaves as if the Einstein Rosen bridge  ends in a bridge-to-nowhere, at least if only local operations are performed. Let us consider it from Alice's viewpoint. She has made a measurement and obtained a result $I.$ She concludes that both Bob's and Charlie's black holes are each in the pure state $|I\ra.$ Therefore there cannot be an ERB connecting them. If she is allowed to classically communicate with Bob and Charlie she can tell them the bad news. This is completely consistent with the mathematical fact that the $GHZ$-state has no entanglement between any two parties.

However the $GHZ$-states are not equivalent to three disconnected bridges-to-nowhere. To see this we may imagine operations in which Alice and Bob cooperate after the measurement.
An example is to allow Alice and Bob   to come together and fuse their black holes into a single black hole called $A\cup B.$ Now $C$ and $A\cup B$ form a bipartite entangled system with $C$ being maximally entangled with $A\cup B.$ Such a bipartite system can be acted on locally at the $A\cup B$ end so that the system is transformed to a TFD state.  Once this has been accomplished then Bob and Charlie can jump in and meet  in the ERB.

Carrying this out after the tensor network has grown  would require a computationally complex operation to shorten the legs of the ERB. It's the  equivalent of undoing Alice's measurement.

There is another way to undo the measurement and restore the smooth ERB connecting Charlie and Bob\footnote{I'm grateful to Patrick Hayden for explaining this to me.}. Suppose that after Alice's measurement,  Daisy observes Alice's memory. If she does her observation in the same basis as Alice's measurement---say the $Z$-basis---then no  change will take place. All that happens is that the combined $A,B,C,D$ system will be in a four-partite $GHZ$-state after Daisy's measurement.

But if  Daisy measures Alice's qubits in the $X$ basis (in the appendix a more general situation is considered) an interesting thing happens. Suppose the outcome of  Daisy's measurement is $X=1$ for all qubits. Then the Charlie-Bob system $B,C$ will be projected back to the TFD state, from which Charlie and Bob can meet in the ERB. 

On the other hand suppose the outcome of  Daisy's message is different, e.g., $X_1=1, \ X_2 = 0, \ X_3=0, \ X_4=1, \ X_5 = 0,....$  In this case $B,C$ will be projected onto a maximally entangled state, but it will not be the TFD. Nevertheless, if  Daisy is allowed to classically communicate the result of her observation to Bob, Bob can carry out a local unitary operation which will restore the TFD. ( Note that Charlie did not have to be involved in the protocol at all.) From this point Bob and Charlie can meet in the ERB.

Both of these ways of restoring the TFD are examples of what Bousso and I called ``un-happening:" a process in which a measurement is undone, and interference between temporarily decohered branches of a wave function is restored \cite{Bousso:2011up}. In principle such un-happenings are possible. The reason we don't usually consider them---even when the apparatus is isolated from the environment---is that apparatuses are generally macroscopic. Once they perform a microscopic measurement they may evolve in a chaotic way. Reversing them is generally computationally very complex and by ordinary standards, impossible: thus the usual assumption that measurement is irreversible.
In the case of a black hole as  apparatus, the growing complexity is represented by the growth of the ERB. Un-happening the measurement involves shortening the ERB which  is  computationally extremely complex \cite{Susskind:2014rva}.

In conclusion: The fact that $GHZ$-entangled black holes are not connected by completely smooth classical Einstein-Rosen bridges is not a problem for ER=EPR; it's  essential to its consistency with quantum mechanics. It ensures that there
is no contradiction between  Daisy's  and Alice's views by having new singular objects which can live in  the ERBs  connecting multipartite entangled systems. The $GHZ$-core is one such object. The properties of these new objects are worth investigating.

 \bn

 \bn

\sc
\section{ERBs as a Resource}\label{S-resource}

Entanglement is often viewed as a resource which is useful for carrying out quantum communication protocols. It stands to reason that if ER=EPR then Einstein-Rosen bridges are also resources for the same purpose. In this section we will see how quantum teleportation can be visualized in terms of ERBs.

For the purposes of this section we may use the traditional description of measurements in terms of the projection postulate:
 the result of acting with an apparatus on a system is to project the system onto a pure state in the measurement basis.  We can dismiss Alice as she will not be needed in what follows.

\begin{figure}[h!]
\begin{center}
\includegraphics[scale=.5]{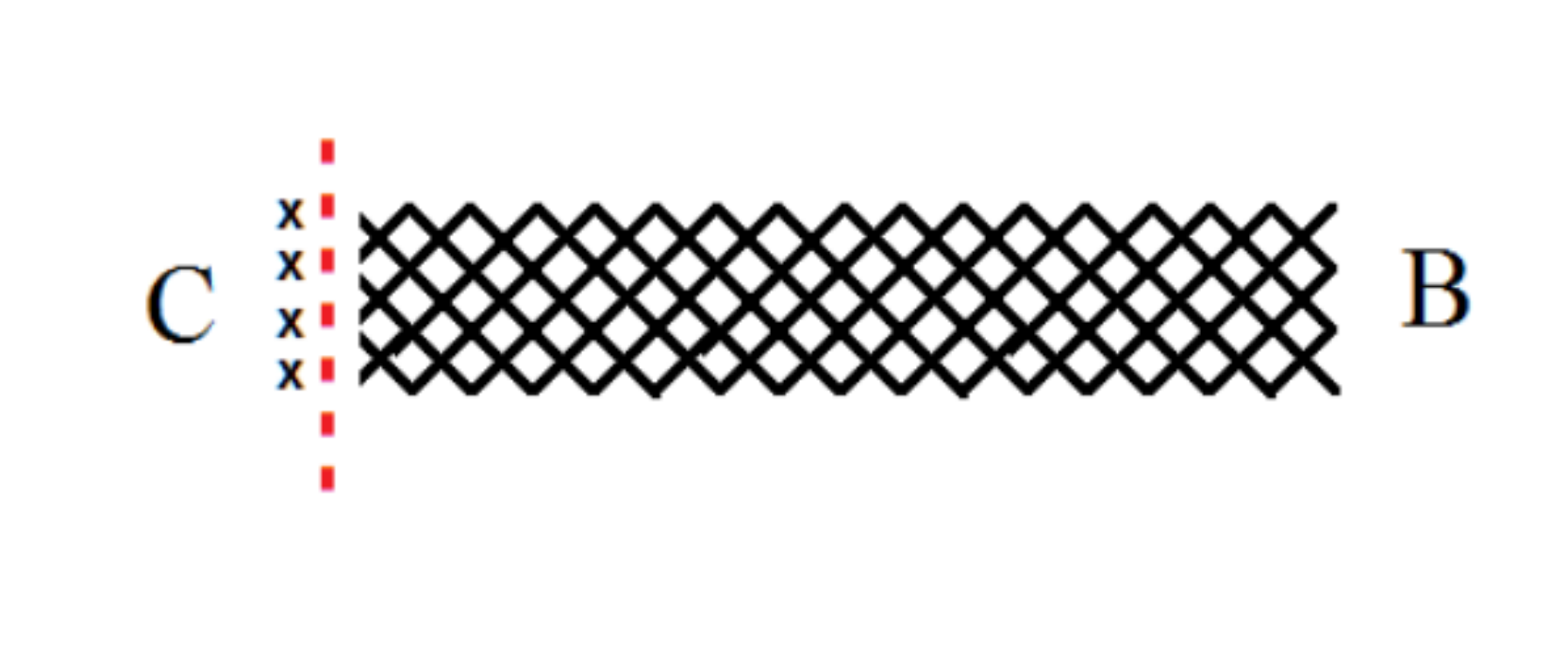}
\caption{Snipping the tensor network representing the ERB connecting Charlie and Bob. }
\label{snip}
\end{center}
\end{figure}

We start with the entangled black hole pair $C,B$ in the state described by the tensor network in the top panel of figure \ref{EVO}, and then allow Charlie to make a complete measurement of his black hole in the computational basis. Charlie's black hole will be projected onto a state $I.$ At the same time Bob's black hole will be projected to the state $U|I\ra = \sum U_{IJ}|J\ra$ where the matrix $U$ represents the tensor network.  Both black holes are projected to  pure states so the outcome state is unentangled. One may say that Charlie has ``snipped" the ERB at his end,
leaving two   bridges-to-nowhere.
   Because the measurement was made at Charlies end, his ERB will be very short.  The snipped ERB is shown in  figure \ref{snip}.

\begin{figure}[h!]
\begin{center}
\includegraphics[scale=.3]{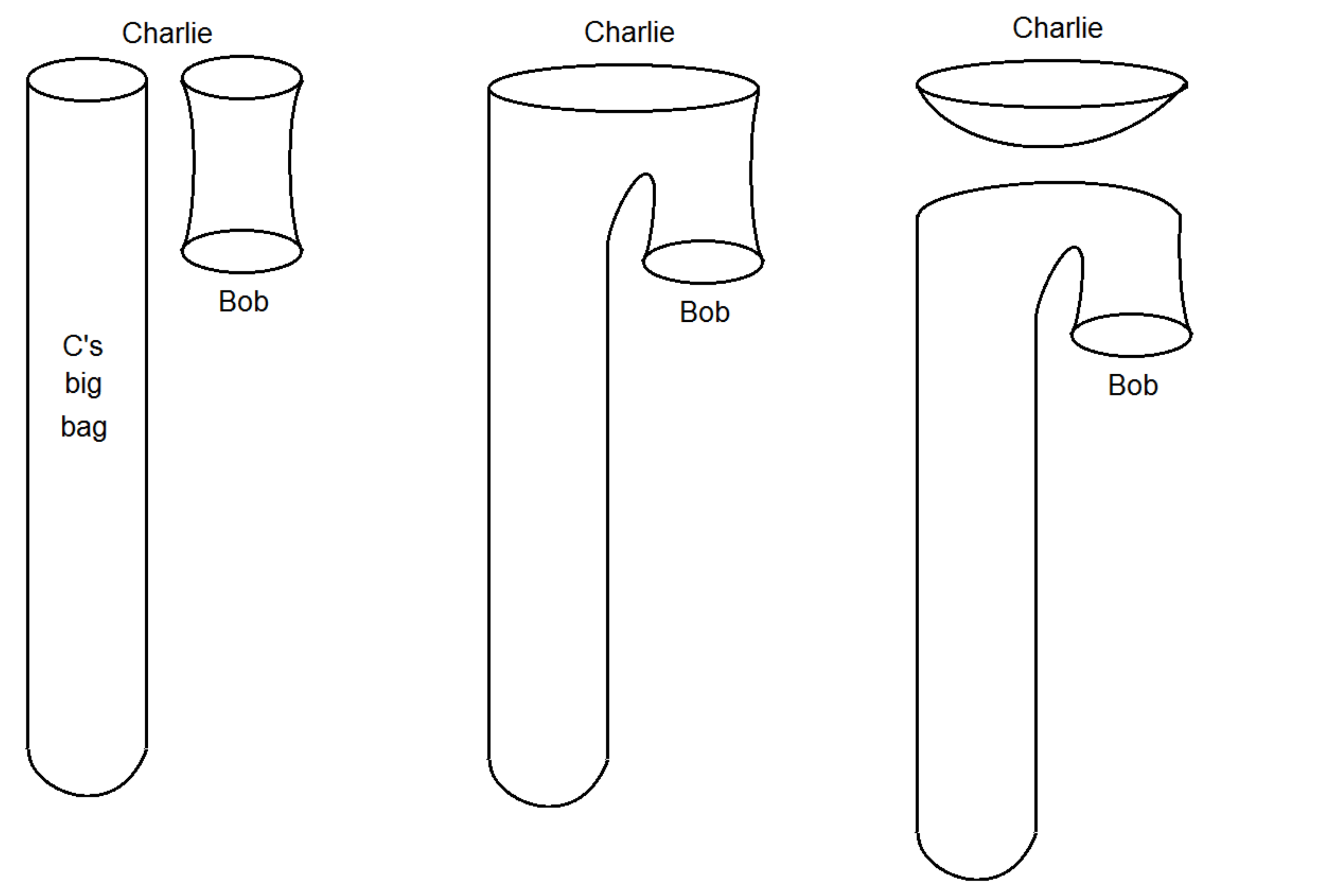}
\caption{The left panel shows the initial state which black holes $B$ and $C$ being entangled, and $C_2$ in a pure state. In the middle panel Charlie has merged $C$ and $C_2.$ In the final panel on the right Charlie has snipped the ERB by making a complete measurement. }
\label{bag}
\end{center}
\end{figure}

Now let's start over with $B$ and $C$ entangled.  A third black hole (for simplicity of the same mass as the other two) will now be added, and given to Charlie so that he now has two black holes in his possession. Charlie's second black hole (call it $C_2$) is in a state,

 \be
 |\psi\ra = \sum_K \psi_K |K\ra.
 \label{pure}
 \ee
Since the state \ref{pure} is pure, $C_2$ is attached to a bridge-to-nowhere. The volume of the the bridge-to-nowhere is determined by the computational complexity of the state of $C_2$ and can be as large as $e^S.$ Let's call it \it Charlie's big bag of complexity.\rm \
The left panel of figure \ref{bag} shows the initial configuration.

Charlie wants to get rid of  the big bag of complexity, so here is what he does: He takes his two black holes, $C$ and $C_2$  and merges them as described in \cite{Susskind:2014moa}. The bag of complexity is fused  with the ERB connecting Charlie and Bob. This is shown in the middle panel of figure \ref{bag}. The geometry of the embedding diagram cannot be drawn in a faithful way but the topology is the important thing.

The final step in Charlie's plan is to wait a scrambling time for the composite $C\cup C_2$ black hole to equilibrate  (the importance of waiting a scrambling time will become clear soon), and then to snip it off (at his end) by making a complete measurement of  $C\cup C_2$  in the computational basis. This cuts the ERB connection between Charlie and Bob, and leaves Bob holding the bag, so to speak.

Of course Bob cannot know what's in the bag unless he knows the outcome of Charlie's measurement;  each outcome will project onto a different state of Bob's black hole. However  Charlie can send the outcome of the measurement in the form of $\sim S$ classical bits of information where $S$ is the entropy of $C\cup C_2$. As we will see, the different outcomes give states which are unitarily related.

 One remarkable thing is that an exponentially complex quantum state has been teleported  to Bob by a process which involved only $S$ classical bits.

What I have described is not a new phenomenon. It is conventional quantum teleportation \cite{Bennett:1992tv} using  an ERB as the entanglement resource. Let us take a look at the mathematics. The initial state is,

\be
\{   \sum_K \psi_ K |K\ra_{C_2}   \} \otimes \{   \sum_{I,J} U_{I,J}|I\ra_{C} \otimes |J\ra_B  \}
\ee
The first factor represents Charlie's unentangled black hole. The complexity resides in the $2^S$ complex coefficients $\psi_K.$  The second factor represents the entangled pair of black holes shared between Bob and Charlie.

I will streamline the notation $|K\ra_{C_2} \otimes |I\ra_{C} \otimes |J\ra_B$ to $|K,I, J \ra$ where the three entries represent states of $C_2, \ C,$ and $B$ in that order.

The next step is to combine the black holes $C$ and $C_2$ into a single system by fusing and then scrambling them leading to a state,

\be
 V \sum_{K,I,J} U_{I,J} \psi_K |K,I,J\ra
\ee
where $ V $ is a unitary scrambling operator acting in the Hilbert space of $C\otimes C_2.$

Finally Charlie's measurement projects  ${C \cup C_2}$ onto an outcome state $ {\la \cal{O}|}$ leaving Bob's black hole in the state,

\be
\sum_{K,I,J}{\la \cal{O}|}V|K,I\ra U_{I,J} \psi_K |J\ra
\ee
The outcome state $ \la \cal{O}|$ may be taken to be a computational state for the qubits comprising $C\cup C_2$ with eigenvalues $\cal{O}.$

Let us fix the outcome state $\la \cal{O}|$ and consider the dependence of $\la {\cal{O}}|V|K,I\ra$ on $I$ and $K.$ This defines a matrix $V_{K,I}$ which implicitly  depends on the eigenvalues of $\cal{O}.$
In general $V_{K,I}$ has no particular special properties. But if the operator $V$  is a random matrix representing the scrambling of $C\cup C_2$ then $V_{K,I}$ will be unitary. The reason is that the scrambled outcome of applying $V$ to the state $\la \cal{O}|$ will be maximally entangled in the degrees of freedom of the product $C\otimes C_2.$ This implies the unitarity of $V_{K,I}.$

Thus the final state of $B$ has the form,

\be
\sum_{K,I,J} V_{K,I} U_{I,J} \psi_K |J\ra =  W|\psi\ra
\ee
with $W$ being a unitary operator parametrically dependent on the outcome.  In other words Bob is left holding a bag which is unitarily related to the original $C_2,$ but the unitary relation depends on the outcome $\cal{O}$ of  Charlie's measurement.

The presence of the scrambler $V$ is essential to getting a unitary operator for $W.$ This is the reason that Charlie waited a scrambling time before making the measurement on $C\cup C_2$.

Of course unless Bob knows the result of Charlie's measuement he cannot know what is in the bag. However, as before, Charlie may send Bob a message consisting of $\sim S$ classical bits which encode the measurement outcome. Following the usual teleportation protocol, once Bob knows the outcome he can apply a unitary operator and bring his black hole to the state \ref{pure}. Thus Charlie's bag of complexity will have been teleported to Bob.

What is new here is not teleportation but the geometric visualization of it in terms of ERBs, and the snipping operation associated with the measurement of one end of an ERB. It is also interesting that a state with large complexity $\sim e^S,$ can be teleported using  a relatively tiny number of classical bits  $\sim S.$
\bn

\sc
\section{Conclusion}\label{S-conclusion}

If there is a lesson to be gained from recent work on quantum gravity it is that geometry and quantum mechanics are so inseparably joined that each may not make sense without the other.

The strategy that I've used for geometrizing a quantum concept  is to apply the concept to a system that either consists of black holes, or can be converted to black holes, and in that way translate the concept into geometric terms\footnote{This is not a new strategy. The earliest example of it was Bekenstein's identification of horizon area  with  entropy.  The most recent example is the relation between the growth of ERBs and the growth of quantum complexity \cite{Susskind:2014rva}\cite{Stanford:2014jda} }. The connection between entanglement and Einstein-Rosen bridges---ER=EPR---is one example. The first part of this paper contains   a heuristic argument that entanglement (EPR) implies wormholes (ER).

Section \ref{S-Measurement and ERBs} is the heart of the paper. It addresses a question that has come up repeatedly since \cite{Maldacena:2013xja} was written: Is the identification of entanglement and ERBs consistent with the standard rules of quantum mechanics?
What we found is  that the geometry of ERBs is not always completely classical; $GHZ$ entanglement gives rise to non-classical obstructions that are important for the consistency of quantum measurements. The non-classical nature of $GHZ$-cores is not a bug; it's a feature necessary for consistency.

Another view of entanglement is that it is a resource that can be used to implement quantum communication.  The last section of the paper  described how quantum teleportation  can be visualized geometrically in terms of the classical fusing of black holes and the quantum snipping of ERBs when measurements take place.

It should be clear that these are just examples, and that we are only scratching the surface of a much deeper relation between quantum mechanics and geometry.

\section*{Acknowledgements}
I thank  Hrant Gharibyan, Patrick Hayden,  Juan Maldacena, and Don Marolf  for helpful discussions about various aspects of $GHZ$-entanglement and Einstein-Rosen bridges. I am also grateful to Douglas Stanford for a discussion that led to the argument of section \ref{S-ER=EPR}.

This work was supported in
part by National Science Foundation grant 0756174 and by a grant from the John Templeton Foundation.
The opinions expressed in this publication are those of the author and do not necessarily
reflect the views of the John Templeton Foundation.

\appendix
\sc
\section{ Generic Measurement by Daisy }

In section \ref{S-Measurement and ERBs} Daisy's measurement of Alice's memory was shown, in certain cases, to undo Alice's measurement and restore the maximal entanglement between black holes $B$ and $C.$ Assuming that the measurement basis for Alice's measurement was $Z$  we found:

\bi
\item If Daisy measures in the $Z$ basis she merely confirms Alice's memory and $B,C$ are left unentangled.
\item If Daisy measures in a conjugate basis such as the $X$ basis, the $B,C$ system will be projected back to a maximally entangled state, although the state depends of the outcome of Daisy's measurement.
\ei

\bn
These two cases are extremes. 
A measurement in a general basis will lie somewhere in between   and result in a $B,C$ entanglement somewhere between the extremes. 

One might wonder about the generic situation: 
 If Daisy measures in a randomly chosen basis what is the expected amount of resulting $B,C$ entanglement? To answer this let $U$ be a  unitary operator acting on Alice's Hilbert space. The two cases we studied are characterized by the matrices $$U =I$$ (the identity operator), and $$U =\prod e^{-i \frac{\pi}{4}Y}$$ where the product is over all qubits in Alice's memory and $Y$ is the second Pauli operator for each qubit. This latter operator is the unitary rotation from the $Z$ basis to the $X$ basis. We will be interested in $B,C$ entanglement that results from more general $U.$

Before Daisy's measurement the tripartite state is $|GHZ\ra$ given by \ref{after measurement}. Let us transform this by acting at Alice's end with $U.$

\be
|\Psi'\ra = U |GHZ\ra =   \sum_{I,J}^{2^K}|I,I,J\ra  U_{JI}
\ee

Next we pick a specific computational state $|O\ra $ in Alice's Hilbert space  and project onto it. The result is some pure state for the $B,C$ system.

\be
|\Psi'\ra_{BC} =  \sum_{I}^{2^K}|I,I\ra  U_{OI}
\ee

From this expression we can calculate the density matrix of Bob's  black hole after tracing over $C.$

\be
\rho_B = \sum_I |U_{OI}|^2 |I\ra \la I|.
\label{rhoB}
\ee

One can measure the degree of entanglement between $B$ and $C$ by computing ${\rm Tr} \rho_B^2.$ For an unentangled pure state this has the value $1$ and for maximally entangled state is the minimal possible value  given by $2^{-K}.$ Using \ref{rhoB},

\be
Tr \rho_B^2 = \sum_I |U_{OI}|^4
\ee

Let us consider average value of this quantity for a randomly chosen $U.$ Using standard formulas for averaging over the unitary group $\suk,$ for large $K$ one finds

\be
Tr \rho_B^2 = 2 \times 2^{-K}
\ee
This is only  a factor of two larger than the minimum which indicates that the entanglement entropy of the $B,C$ system is within about one bit of being  maximal.

We can interpret this in one of two ways. The first is that if Daisy makes a measurement of Alice's memory in a random basis the $B,C$ system will be projected onto an almost maximally entangled state.

The second interpretation is that if the dynamics of Alice's memory system is chaotic, and if Daisy waits for the memory to scramble, then Daisy's measurement can be done in the computational basis and the resulting $B,C$ system will almost always be within one bit of being maximally entangled.

In either case the implication is that if Daisy is permitted to classically transmit the outcome to Bob, he can carry out a local unitary operation which will restore the TFD, after which Bob and Charlie's signals can meet in the ERB.

Nothing in this argument should be interpreted as implying that it is easy to undo Alice's measurement.  It is true that Daisy does not have to fine-tune the measurement basis, and it is not especially difficult for her to classically communicate the outcome to Bob. The difficulty  is in the next step; the unitary operator that Bob has to apply will generically be very computationally  complex.

\end{document}